\pdfoutput=1
\documentclass[9pt, twocolumn]{article}

\usepackage{graphicx}
\usepackage{amssymb}
\usepackage{epstopdf}
\usepackage{todonotes}
\usepackage{hyperref}
\usepackage{color}
\usepackage{setspace}
\usepackage{amsmath}
\usepackage{amsthm}
\usepackage{dsfont}
\usepackage{todonotes}
\usepackage{accents}
\usepackage{epstopdf}
\usepackage{bm}
\usepackage{todonotes}
\usepackage{float}

\newcommand{\w}{\mathbf{w}}

\newcommand{\uu}{u}

\newcommand{\F}{{F}}
\newcommand{\hF}{\hat{F}}
\newcommand{\tF}{{F'}}
\newcommand{\htF}{\hat{{F'}}}
\newcommand{\he}{\mathbf{\hat{n}}}
\newcommand{\ee}{{\mathbf{n}}}

\newcommand{\GG}{\mathbf{G}_{F}}
\newcommand{\tGG}{{\mathbf{G}'}_{{F'}}}

\newcommand{\dw}{\dot{\bm{\omega}}_T}
\newcommand{\dta}{\bm{\delta}}
\newcommand{\hdw}{\dot{\hat{\bm{\omega}}}_T}
\newcommand{\uy}{y}
\newcommand{\uh}{h}
\newcommand{\tu}{{u'}}
\newcommand{\ty}{{y'}}
\newcommand{\tW}{{W'}}

\newcommand{\bg}{{\bar{\gamma}}}

\title{Fundamental bounds on learning performance in neural circuits}

\begin{document}
\author{Dhruva V. Raman, Timothy O'Leary}
\date{August 2, 2018}
\maketitle



\begin{abstract}
How does the size of a neural circuit influence its learning performance? Intuitively, we expect the learning capacity of a neural circuit to grow with the number of neurons and synapses. Larger brains tend to be found in species with higher cognitive function and learning ability. Similarly, adding connections and units to artificial neural networks can allow them to solve more complex tasks. However, we show that in a biologically relevant setting where synapses introduce an unavoidable amount of noise, there is an optimal size of network for a given task. Beneath this optimal size, our analysis shows how adding apparently redundant neurons and connections can make tasks more learnable. Therefore large neural circuits can either devote connectivity to generating complex behaviors, or exploit this connectivity to achieve faster and more precise learning of simpler behaviors. Above the optimal network size, the addition of neurons and synaptic connections starts to impede learning performance. This suggests that overall brain size may be constrained by the need to learn efficiently with unreliable synapses, and may explain why some neurological learning deficits are associated with hyperconnectivity. Our analysis is independent of specific learning rules and uncovers fundamental relationships between learning rate, task performance, network size and intrinsic noise in neural circuits.
\end{abstract}





\section*{Introduction}
In the brain, computations are distributed across circuits that can include many millions of neurons and synaptic connections. Maintaining a large nervous system is expensive energetically and reproductively \cite{laughlin1998metabolic,tomasi2013energetic,attwell2001energy}, suggesting that the cost of additional neurons is balanced by an increased capacity to learn and process information.

Empirically, a `bigger is better' hypothesis is supported by the correlation of brain size with higher cognitive function and learning capacity across animal species \cite{reader2002social, sol2005big, joffe1997visual}. Within and across species, positive correlations exist between the volume of a specific brain region and the behavioural importance of tasks related to the brain region in question \cite{maguire2000navigation, gaser2003brain, black1990learning}. 
These observations make sense from a theoretical perspective because larger artificial neural networks can be trained to solve more challenging 
computational tasks than smaller networks \cite{lawrence1998size, krizhevsky2012imagenet, huang2003learning, takiyama2016maximization, takiyama2012maximization, saxe2013exact}.

However, biologically it is not clear that there is always a computational advantage to having more neurons and synapses engaged in learning and solving a task. During learning, larger networks face the problem of tuning a greater number of synapses using limited and potentially corrupted information on task performance \cite{seung2003learning, werfel2004learning}. Moreover, no biological component is perfect, so unavoidable noise arising from the molecular machinery in individual synapses might sum unfavourably as the size of a network grows. Intriguingly, a number of well-studied neurodevelopmental disorders exhibit cortical hyperconnectivity at the same time as learning deficits \cite{contractor2015altered, rinaldi2008hyper, casanova2006minicolumnar, amaral2008neuroanatomy}. It is therefore a fundamental question whether learning capacity can grow indefinitely with the number of neurons and synapses in a neural circuit, or whether there is some law of diminishing returns that eventually leads to a decrease in performance beyond a certain network size.

We address this question with a general mathematical analysis of learning performance in neural circuits that is independent of specific learning rules and circuit architectures. 
Given some fixed task, we show analytically that expected learning rate and steady-state performance is higher in larger networks than smaller networks when there is no intrinsic noise in the connections. The analysis reveals how connections can be added to a network in a principled way to reduce the difficulty of learning a task. This gain in overall learning performance is accompanied by slower per-synapse rate of change, predicting that synaptic turnover rates should vary across brain areas according to the number of connections involved in a task, and the typical task complexity.

If each synaptic connection is intrinsically noisy we show that there is an optimal network size for a given task. Above the optimal network size, adding neurons and connections degrades learning and steady-state performance. This reveals an important disparity between synapses in artificial neural networks, which are not subject to unavoidable intrinsic noise, and those in biology, which are necessarily subject to fluctuations at the molecular level \cite{loewenstein2015predicting,ziv2017synaptic,minerbi2009long,puro1977synapse}.

For networks that are beneath the optimal size, it turns out to be advantageous to add apparently redundant neurons and connections. We show how additional synaptic pathways reduce the impact of imperfections in learning rules and uncertainty in the task error. This provides a potential theoretical explanation for recent, counterintuitive experimental observations in mammalian cortex \cite{bloss2018single, bartol2015nanoconnectomic}, which show that neurons frequently make multiple, redundant synaptic connections to the same postsynaptic cell. A non-obvious consequence of this result is that the size of a neural circuit can either reflect the complexity of a fixed task, or instead deliver greater learning performance on simpler, arbitrary tasks.

\section*{Results}

\subsection*{Modelling the effect of network size on learning}

\begin{figure}[h]
\centering
\includegraphics[width=1.0\linewidth]{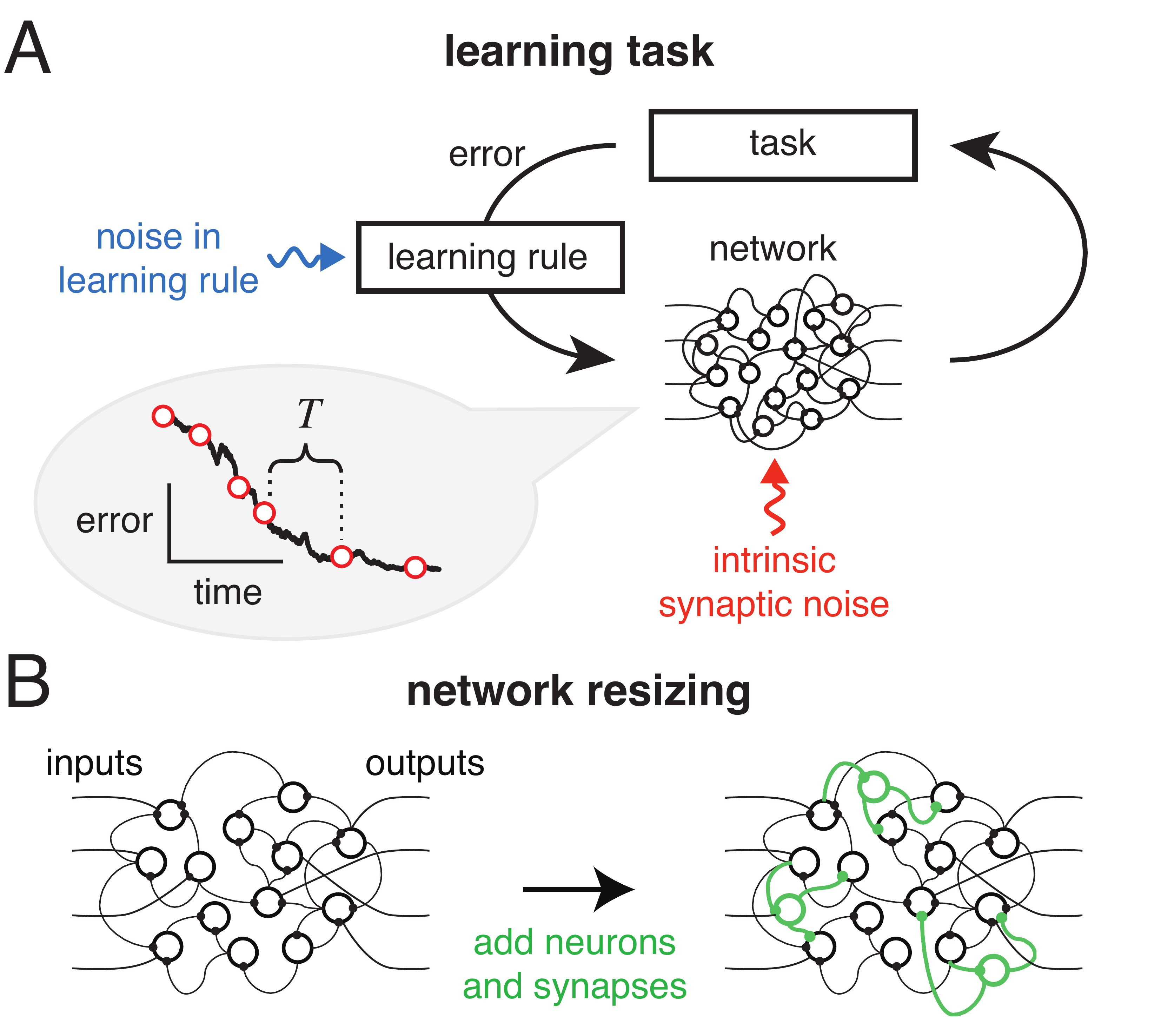}
\caption{\label{fig:1} \textbf{A}: Schematic of learning in a neural network. Information on task error is received by a learning rule which converts this information into synaptic changes that decrease task error. Biologically, the learning rule faces several challenges: it will be subject to noise and perturbations (blue arrow), and the synapses themselves may suffer from intrinsic noise (red arrow). Error information will only acquired be intermittently, as shown in the learning curve on the left, where $T$ specifies the intermittency of feedback (see main text).
\textbf{B}: We analyse the effect of network size on learning performance by adding redundant neurons and synapses (green) to an existing network.
}
\end{figure}

Our goal is to analyse how network size affects learning and task performance in a general setting depicted in Figure \ref{fig:1}, which is independent of specific tasks, network architectures and learning rules. We assume that there is some error signal that is fed back to the network via a learning rule that adjusts synaptic weights. We also assume that the error signal is limited both by noise and by a finite sampling rate quantified by some time interval $T$ (Figure \ref{fig:1}A). In addition to the noise in the learning rule, we also consider noise that is independently distributed across synapses (`intrinsic synaptic noise'). This models molecular noise in signalling and structural apparatus in a biological synapse that is uncorrelated with learning processes, and with changes in other synapses. Network size is adjusted by adding synapses and neurons (Figure \ref{fig:1}B).

\begin{figure}[p]
\centering
\includegraphics[width=1\linewidth]{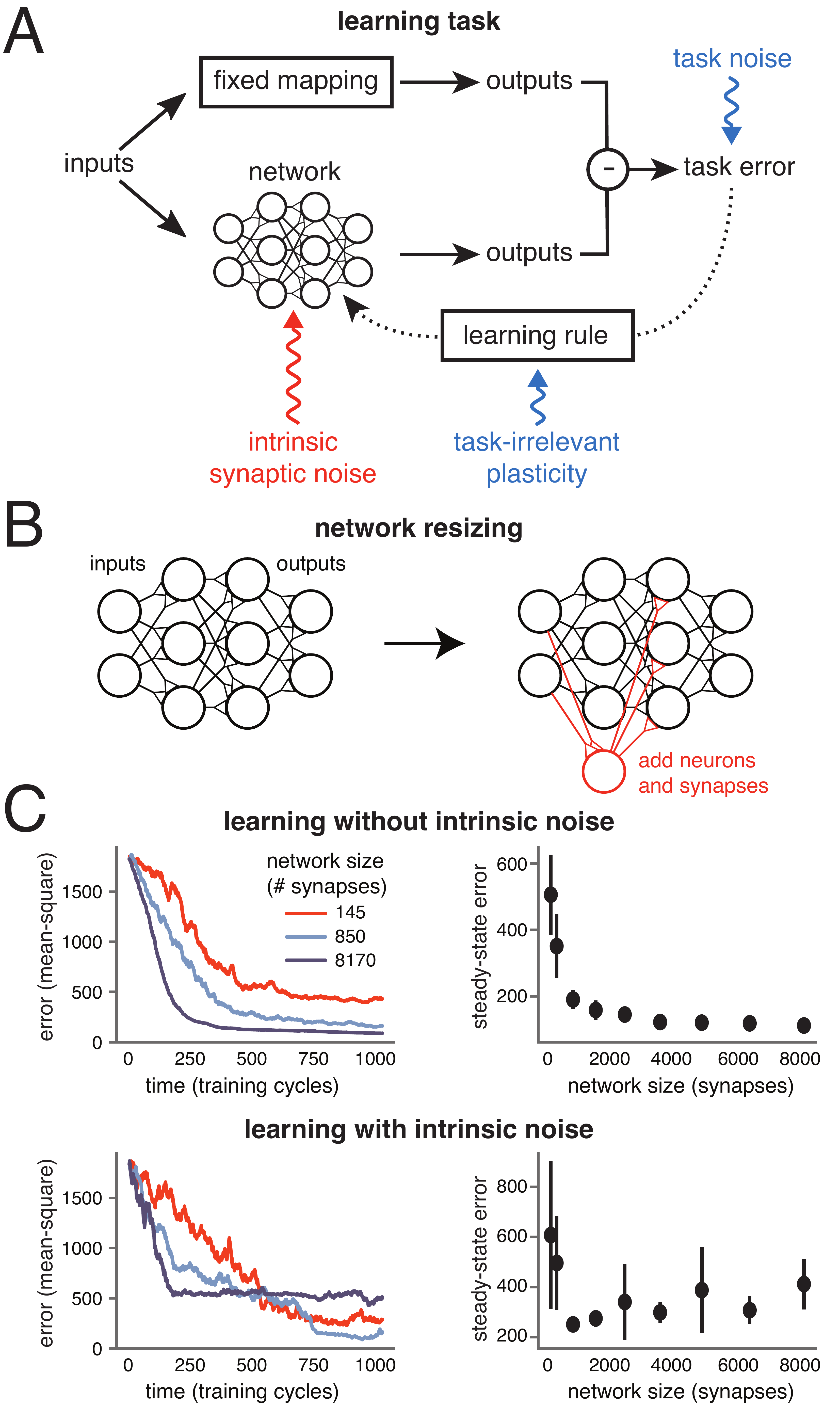}
\caption{
\label{fig:2}
\textbf{A} Learning task: neuronal networks are trained to learn an input-output mapping using feedback error and a gradient-based learning rule to adjust synaptic strengths. The feedback is corrupted with tunable levels of noise (blue), reflecting imperfect sensory feedback, imperfect learning rules, and task-irrelevant changes in synaptic strengths. Synaptic strengths are additionally subject to independent internal noise (red), reflecting their inherent unreliability.
\textbf{B}: Network size is increased by adding neurons and synapses to inner layers. 
\textbf{C}: Three differently sized networks are trained on the same task, with the same noise-corrupted learning rule. (Right) Mean task error after $1000$ training cycles, computed over $12$ simulations. Error bars depict $\pm 1$ standard deviation  from the mean. (Left) task error over time for a single simulation of each network.
\textbf{D}:  As for (C) but each synapse is subject to internal independent noise fluctuations in addition to noise in the learning rule.
}
\end{figure}

Before dealing with the general case, we motivate the analysis with simulations of standard nonlinear feedforward neural network models that we trained to compute arbitrary input-output mappings, as shown in Figure  \ref{fig:2}A. We simulate learning in networks using the so-called student-teacher framework, where we generate a fixed input-output mapping using a nonlinear feedforward `teacher' network with randomly chosen fixed synaptic weights. Importantly, teacher networks can generate arbitrarily complex input-output mappings with the convenient property of being learnable by a `student' network with the same underlying connection topology. We added neurons and connections to internal layers to generate student networks of increasing size (Figure \ref{fig:2}B). This scales network size while ensuring that all networks can solve the same task.

Learning was simulated by modifying synapses with noise-corrupted gradient descent to mimic an imperfect biological learning rule. We emphasize that do not assume that learning in a biological network occurs by explicit gradient descent. However any error-based learning rule must induce synaptic changes that approximate gradient descent, as we show below (Equation \ref{eq:intCorr}).

The phenomenon we wish to understand in detail is shown in Figures \ref{fig:2}C and D. We trained networks of varying sizes on the same task, with the same amount of learning rule noise. Larger networks learn more quickly and to a higher steady state performance than smaller networks when there is no intrinsic noise in the synapses Figure \ref{fig:2}C. This is surprising because the only difference between the smallest network and larger networks is the addition of redundant synapses and neurons, and the task is designed so that all networks can learn it perfectly in principle. Moreover, as shown in Figure  \ref{fig:2}D, a small amount of internal noise in each of the connections of the student networks results in a non-monotonic relationship between performance and network size. Beyond a certain size, both learning and asymptotic performance start to worsen.

The simulations in Figure  \ref{fig:2} provide evidence of an underlying relationship between learning rate, task performance, network size and intrinsic noise. To understand these observations in a rigorous and general way, we mathematically analysed how network size and imperfections in feedback learning rules impact learning in a general case. 

\subsection*{Learning rate and task difficulty}
We first establish a general relationship between learning rate and key properties of error-based learning rules. We define task error as a smooth function $\F[\w]$ which decreases with increasing levels of performance, and which depends on the vector $\w$ of $N$ synaptic weights in a network. We assume that learning rules use some (potentially imperfect) estimate of this error to adjust synaptic weights.

Biologically, it is reasonable to assume that learning-related synaptic changes occur due to old information. For example, a task-related reward or punishment may be supplied only at the end of a task, which itself takes time to execute. Similarly, even if error feedback is ongoing in time, there will always be some biochemically induced delay between acquisition of this error signal and its integration into plastic changes at each synapse.

\begin{figure}[h]
\centering
\includegraphics[width=1.0\linewidth]{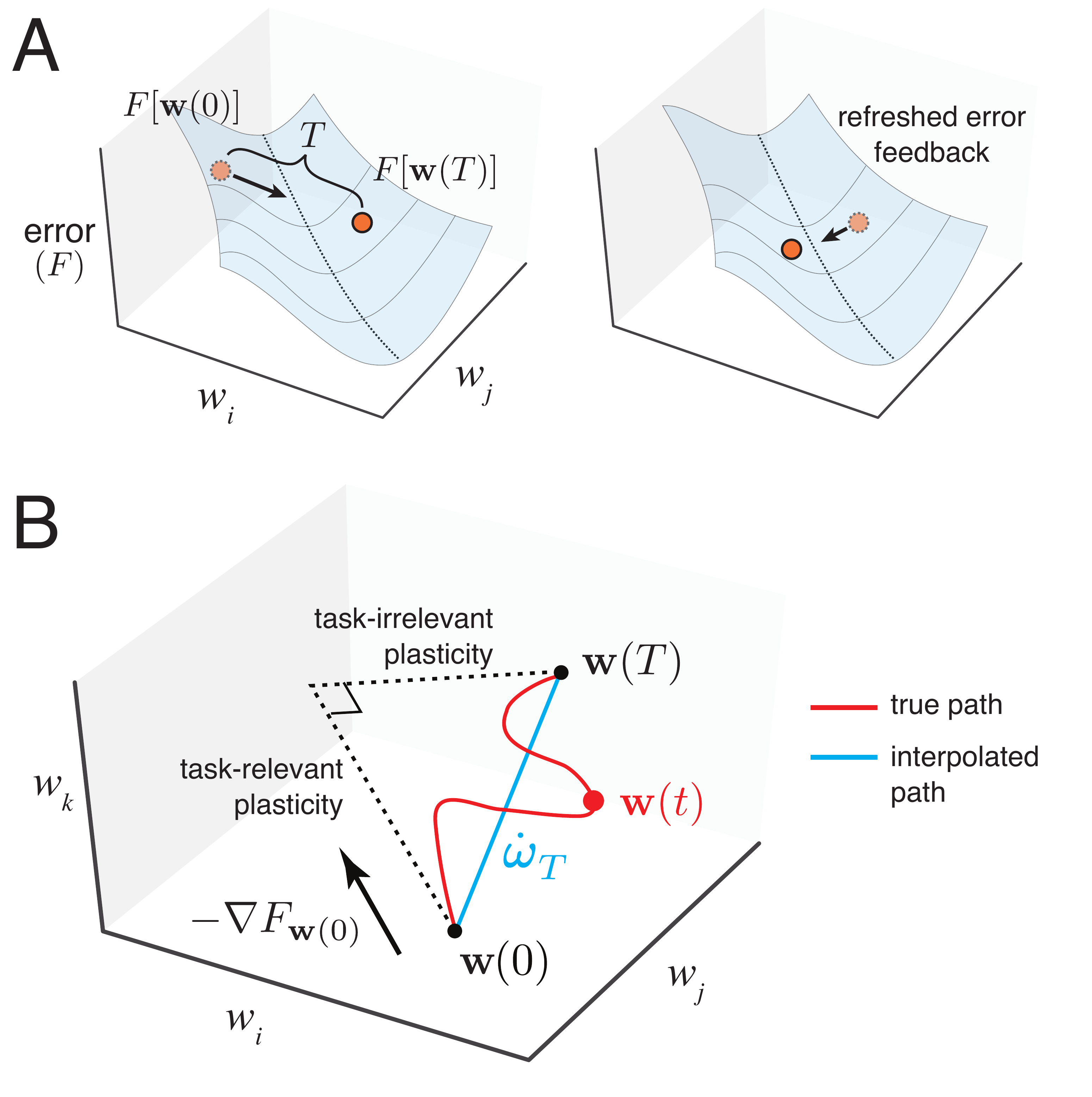} 
\caption{\label{fig:3} 
Geometry of error-based learning in arbitrary networks. \textbf{A}: Schematic of task error as a function of (two) synaptic weights. Learning rule receives and processes task-relevant feedback to provide direction for each synapse to move in weight space. Direction must correlate with steepest descent direction, resulting in initial improvement of task error. If no new feedback is received over a long time period $T$, this initially good direction may eventually go uphill, thus becoming bad. Frequent error feedback, a less `curvy' error surface, and a good correlation with the initial steepest descent direction make learning faster. Local task difficulty captures these factors. 
\textbf{B}: Schematic of changes in three weights over interval $[0,T]$. The true weight trajectory $\w(t)$ over time (red line) is summarised by an interpolated, linear trajectory $\dw$ (blue line) between $\w(0)$ and $\w(T)$. We can decompose this interpolated trajectory into task-relevant and task-irrelevant plasticity components.  The former is the initial steepest descent direction $-\nabla \F[\w(0)]$, and the latter is the remaining orthogonal component.
}
\end{figure}

Thus, there will be a maximum rate at which task error information can be transmitted to synapses during learning, which for mathematical convenience can be lumped into discrete timepoints. Suppose feedback on task error occurs at timepoints $0$ and $T$, but not in between, for some $T>0$ (Figure \ref{fig:3}A). If the network learned over the interval $[0, T]$ then $\F[\w(T)] - \F[\w(0)] < 0$ by definition. We quantify learning rate during this interval as the value of $k$ such that
\begin{align*}
\F[\w(T)] = (1 - {kT }) \F[\w(0)],
\end{align*}

with $k<\frac{1}{T}$. A larger positive value of $k$ implies a faster rate of learning. We can write the total change in error over the interval $T$ (see Figure \ref{fig:3}A) as:
\begin{align}
\F[\w(T)] - \F[\w(0)] &= \int_0^{T} \langle \nabla \F[\w(t)], \dot{\w}(t) \rangle \ dt \nonumber \\
&= T \ \mathbb{E}_t[\langle \nabla \F[\w(t)], \dot{\w}(t) \rangle], \label{eq:intCorr}
\end{align}

where expectation is taken across a uniform distribution of timepoints in $[0,T]$, dots denote time derivatives and angle brackets denote the (standard) inner product. Equation (\ref{eq:intCorr}) shows that synaptic changes, on average, must anticorrelate with the gradient for learning to occur. We can thus decompose net learning rate during the interval $T$ into contributions as follows (further details in {\em SI section `Learning rate and local task difficulty'}):
\small{\begin{align}
\begin{split}
k  = \underbrace{
   \tfrac{ -\|\nabla \F[\w(0)]\|_2}{\F[\w(0)]} 
  }_{\substack{\text{gradient} \\ \text{strength}}} 
 \left[  \overbrace{ \langle  \dw, \nabla \hF[\w(0)] \rangle }^{\substack{\text{contribution} \\ \text{from gradient}}} + \underbrace{ \GG[ \hdw] \| \dw \|^2_2 T}_{\substack{\text{contribution} \\ \text{from curvature}}}  \right]  \\ + \mathcal{O}(T ^2)
 \end{split}
 \label{eq:2orderStd}
\end{align}}
where
\begin{align*}
\GG[ \hdw] &:= \frac{1}{2 \|\nabla \F[\w(0)]\|_2}  \left\langle \hdw, \nabla^2 \F[\w(0)] \hdw \right\rangle. 
\end{align*}

Hats indicate unit length normalized vectors (i.e. $\hat{x} = \frac{x}{\|x\|_2}$) and $\dw$ denotes the average synaptic change over the time interval $[0,T]$, normalised by $T$:
\begin{align}
 \dw = \frac{\w(T)  - \w(0)}{T}. \label{eq:constVel} 
 \end{align}

Note that we have made no assumptions on the size of $T$, so the $\mathcal{O}(T^2)$ term in \eqref{eq:2orderStd} is not necessarily small. Nonetheless, we can gain useful insight for how error surface geometry affects learning by examining the other terms on the right hand side of \eqref{eq:2orderStd}. The gradient strength scales the overall learning rate. Inside the brackets, the curvature term (which can change sign and magnitude during learning) can compete with the gradient term to slow down (or reverse) learning.

Informally, the curvature term in \eqref{eq:2orderStd} therefore controls the learning `difficulty' at each stage of learning. As we will show, this term can be tuned by changing the number of neurons and synaptic connections in the network. 

The learning rate, $k$, is likely to remain positive during learning if the gradient direction changes gradually as the error surface is traversed (i.e. the error surface is almost linear). In this case a high rate of plasticity - due to a high gain between feedback error and synaptic change - will result in a high learning rate. However, if the descent direction changes rapidly due to the curvature of the error surface (i.e. the surface is crinkled up), then correlation with $- \nabla \F[\w(0)]$ becomes a weaker predictor of learning over the entire time interval $T$. Effective learning therefore involves balancing error surface curvature and per-synapse rate of plasticity. This is illustrated in Figure \ref{fig:3}A, where the length of the leaps along the error surface indicate the rate of plasticity.


We next decompose the contributions to the overall synaptic change during a learning increment. First we assume that synapses are perfectly reliable, with no intrinsic noise fluctuations affecting their strengths. In this case, we can decompose $\dw$ into two components that are parallel and perpendicular to the gradient at time $0$, when error information was supplied to the network (see Figure \ref{fig:3}B):
\begin{align*}
\dw = -\gamma_1 \nabla \hF[\w(0)]  + \gamma_2 \he_{2}.
\end{align*}

where $\gamma_1$ is the component of synaptic change that projects onto the error gradient direction and $\gamma_2$ is the component perpendicular to the gradient direction, with $\he_{2}$ denoting the unit vector in this direction. We call these two components, $\gamma_1$ and $\gamma_2$, the \textbf{task-relevant plasticity} and \textbf{task-irrelevant plasticity} respectively. 

Note that a learning rule could theoretically induce task-relevant synaptic changes in a direction that is not parallel to the gradient, $\nabla \F[\w(0)]$, if information on the Hessian $\nabla^2 \F[\w(0)]$ were available. However, as mentioned previously we are assuming that such information is not available biologically and the best the network can do is follow the gradient. In fact, the results of the paper can be generalised so as to eliminate this assumption (see {\em SI section `Task-relevant Plasticity'}) but this complicates the presentation without adding insight.

There are several sources of task-irrelevant plasticity. Firstly, there can be inherent imperfections in the learning rule: information on task error may be imperfectly acquired and transmitted through the nervous system. Secondly, as we have emphasized above, the process of integrating feedback error and converting it into synaptic changes takes time. Therefore any learning rule will be using outdated information on task performance, implying that the gradient information will have error in general, unless it is constant for a task. This is illustrated in Figure \ref{fig:3}A, where we see that during learning, the local information used to modify synapses leads to a network overshooting local minima in the error surface. Thirdly, in a general biological setting, synapses will be involved in multiple task-irrelevant plasticity processes that contribute to $\gamma_2$ (Figure 1A). For instance, the learning of additional, unrelated tasks may induce concurrent synaptic changes; so too could other ongoing cellular processes such as homeostatic plasticity. The common feature of all these components of task-irrelevant plasticity is that they are correlated across the network, but uncorrelated with learning the task.

We now consider the impact of intrinsic noise in the synapses themselves. Synapses are subject to continuous fluctuations due to turnover of receptors and signalling components. Some component of this will be uncorrelated across synapses so we can model these sources of noise as additional, independent white-noise fluctuations at each synapse with some total (per-synapse) variance $\gamma_3 T$ over the interval $[0, T]$. Because this noise is independent across synapses the total variance in the network will scale with the number of synapses. This gives a new expression for synaptic weight change:
\begin{align}
\w( T) - \w(0) &=  -\gamma_1 T \nabla \hF[\w(0)]  + \gamma_2 T \he_2  + 
\gamma_3 \sqrt{T}\sqrt{N} \he_3 \nonumber \\
&= T \left( -\gamma_1  \nabla \hF[\w(0)]  + \gamma_2  \he_2  +
\gamma_3 \sqrt{\frac{N}{T}} \he_3. \right) \label{eq:weightDecomp}
\end{align}

Note that $\gamma_3$ describes the average degree of intrinsic noise per synapse, whereas $\gamma_1$ and $\gamma_2$ describe components of synaptic change over the entire network. This in turn gives the following expression for the average weight velocity over the learning interval:
\begin{align}
 \dw = -\gamma_1 \nabla \hF[\w(0)]  + \gamma_2  \he_{2}  + 
\gamma_3 \sqrt{\frac{N}{T }}
\he_{3}. \label{eq:decompVelocity}
\end{align}

If we assume that each component is independent (see {\em SI section `Decomposition of local task difficulty'} for justification) we can also write down the magnitude of total synaptic rate of change in a convenient form:
\begin{align}
\|\dw\|^2_2 = {\gamma_1^2 + \gamma_2^2 + \gamma_3^2\frac{N}{T }}. \label{eq:expGamma}
\end{align}

Expressions \eqref{eq:decompVelocity} and \eqref{eq:expGamma} allow us to rewrite \eqref{eq:2orderStd}:
\begin{align} 
\begin{split}
\mathbb{E}(k) =  \frac{ -\|\nabla \F[\w(0)]\|_2}{\F[\w(0)]}  \Big[ {-\gamma_1} &+  \GG[\hdw] \|\dw\|^2_2T  \Big]  \\  &+ \mathcal{O}(T ^2). 
\end{split} \label{eq:simpleK}
\end{align}

\begin{figure}[t]
\centering
\includegraphics[width=0.8\linewidth]{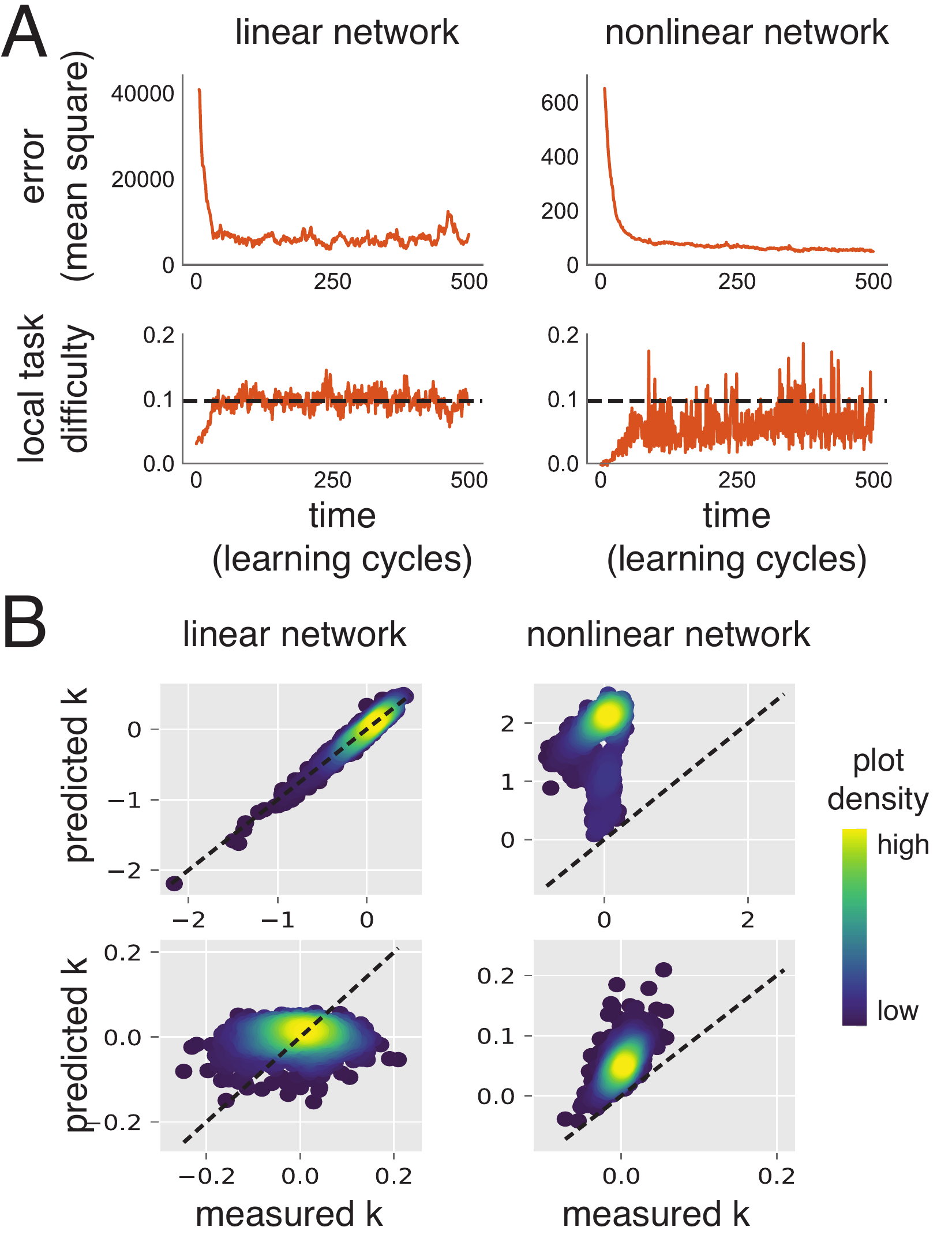} 
\caption{\label{fig:4} Numerical validation of learning rate calculations in simulated neural networks. \textbf{A}: 
Local task difficulty and MSE over time for a linear network (left) with quadratic error function, and a nonlinear network (right). Local task difficulty is low when the networks are in an untrained state. As performance improves, it rises, until reaching some steady state level (black dotted lines). We can predict this steady state \emph{a priori}, exactly for the quadratic error, and conservatively for the nonlinear error, using \eqref{eq:speedLimit}. Both networks are trained using a corrupted learning rule ($\bg=[0.2,1,0]$, $T=2$, see Methods). Network sizes are $200$ synaptic weights (linear) and $220$ synaptic weights (nonlinear, one hidden layer).
\textbf{B}: We use the same linear (left) and nonlinear (right) networks as in A. We compare predicted value of the learning rate $k_{\text{pred}}$ (using \eqref{eq:simpleK} and $ \gamma = \bg$) with the actual value, under low-noise (top , $\bg = [1,0.05,0.05]$) and high-noise (bottom, $\bg = [0.2,0.5,0.1]$) conditions. 
Dotted lines represent $k = k_{\text{pred}}$. Density plots of $\{k, k_{\text{pred}}\}$ are shown. Two sources of discrepancy exist. First, $k_{\text{pred}}$ is calculated from the mean values $\bg$ (see Methods). Transient correlations between task-irrelevant sources of plasticity and the gradient lead to unbiased fluctuations of $\gamma$ around $\bg$. This is the only source of discrepancy in the linear case (left). Thus the density distributes equally on either side of the dotted line. 
 In the nonlinear case (right), there is an unknown, nonzero $\mathcal{O}(T^2)$ term (\eqref{eq:simpleK}) unaccounted for in calculation of $k_{\text{pred}}$. This term almost always decreases learning rate, as $k_{\text{pred}}$ now consistently overestimates $k$. Thus predicted steady state local task difficulty (e.g. dotted line, bottom-right panel of (A)) is consistently an overestimate. 
}
\end{figure} 

If we fix values of the $ \gamma_i$ and $T$, we see from \eqref{eq:simpleK} that $\GG$ controls the learning rate of a network: a higher value of $\GG[\hdw]$ leads to slower or negative learning. For this reason, we refer to $\GG$ as the \textbf{local task difficulty}. Again, the $\mathcal{O}(T^2)$ term may not be small. However, as mentioned, it is reasonable to assume this term cannot be controlled by the learning rule because it depends on higher order derivatives of $\F$ that synapses are unlikely to be able to compute. Therefore we can reasonably say that learning requires the first term of \eqref{eq:simpleK} to be negative, and ceases to occur when this term is zero. This implies:
\begin{align}
\GG[\hdw] \leq \frac{\gamma_1}{T  \Big( \gamma_1^2 + \gamma_2^2 + \gamma_3^2\frac{N}{T }\Big) }. \label{eq:speedLimit}
\end{align}

This inequality relates the intrinsic `learnability' of a task (local task difficulty, $\GG$), the rate of information on task error ($T$), the quality of the learning rule (relative magnitudes of $\gamma_1$ and $\gamma_2$), the network size ($N$), and the intrinsic noisiness of synapses ($\gamma_3$).   

If inequality \eqref{eq:speedLimit} is broken, then learning stops entirely. 
At some point in learning, this breakage is inevitable: as $\F[\w]$ approaches a local minimum, the gradient $\nabla \F[\w]$ approaches zero, and the Hessian $\nabla^2 \F[\w]$ is guaranteed positive semidefinite. At a precise minimum of error, $\GG$ becomes unbounded. This means that for a nonzero $T$, cessation of learning is preceded by an increase in local task difficulty, and learning stops just as inequality (\ref{eq:speedLimit}) above is broken.

To validate our analysis we numerically computed the quantities in \eqref{eq:simpleK} in simulations (Figure \ref{fig:4}). In the case of a linear network with quadratic error the $\mathcal{O}(T^2)$ terms disappear, allowing us to verify that equality in \eqref{eq:speedLimit} indeed predicts the steady-state value of $\GG$. This agreement is confirmed in Figure \ref{fig:4}A.

For more general error functions, we have observed that \eqref{eq:speedLimit} is always conservative in numerical simulations: learning stops before local task difficulty reaches the critical value value, implying that the $\mathcal{O}(T^2)$ term of \eqref{eq:simpleK} is usually negative. This is demonstrated in Figures \ref{fig:4}A and \ref{fig:4}B.

In summary, we have shown that local task difficulty $\GG$ determines the learning rate as well as the steady state learning performance of a network.

\subsection*{Local task difficulty as a function of network size}

We next show precisely how network size influences the local task difficulty, and thus learning rate and steady state performance when other factors such as noise and the task itself remain the same.

Recall that $\ee_{2}$ represents the direction in which synapses are perturbed due to error in the learning rule and other task-irrelevant changes that affect all synapses. Meanwhile $\ee_{3}$ represents the direction of weight change due to intrinsic white-noise fluctuations at each synapse. For arbitrary tasks, networks and learning trajectories we can model these terms as coming from mutually independent probability distributions that are independent of task error $\F[\w]$ and its derivatives. Thus we assume $\mathbb{E}[\ee_{2}^T \ee_{3}] = 0$, for any $i$, which allows us to write an expression for expected local task difficulty:
\small{\begin{align}
\mathbb{E}\big[\GG[\hdw] \big] = \gamma_1^2 \GG^1[\w(0)] +  \frac{Tr(\nabla^2 \F[\w(0)])}{2 \| \nabla \F[{\w(0)}] \|_2 } \left[  \frac{\gamma_2^2  }{N} +    \frac{\gamma_3^2  }{T} \right], \label{eq:gdeDecomp}
\end{align}}
where
\small{
\begin{align}
&\GG^1[\w(0)] = \frac{1}{2 \| \nabla \F[\w(0)] \|_2}  \Big\langle {\nabla \hF[\w(0)]}, \nabla^2 \F[\w(0)]  {\nabla \hF[\w(0)]} \Big\rangle.\label{eq:GG1}
\end{align}}

This new expression for the local task difficulty explicitly incorporates $N$, the number of synaptic weights. So too does the learning rate \eqref{eq:simpleK}, as we see by substituting into it the expanded forms of $ \mathbb{E}[\|\dw\|^2_2]$ and $\mathbb{E}\big[\GG[\hdw] \big]$.

We can gain intuition into how \eqref{eq:gdeDecomp} is derived without going through additional technical details ({\em SI section `Decomposition of local task difficulty'}). Suppose that the weights were perturbed by a randomly chosen direction $\ee$ over the time interval $[0, T]$. This gives:
\begin{align}
\begin{split}
&\F[\w(T)] = \F[\w(0) + \ee] = \\ \ &F[\w(0)] +  \langle \nabla \F[\w(0)], \ee \rangle + \langle \ee, \nabla^2 \F[\w(0)] \ee \rangle + \mathcal{O}(\|\ee\|_2^2) \label{eq:hessianNoiseExp}
\end{split}
\end{align}

If the direction $\ee$ is drawn independently of task error and its derivatives, then
\begin{align*}
& \mathbb{E}[ \langle \nabla \F[\w(0)], \ee \rangle] = 0.
\end{align*}

Therefore it is the quadratic term of \eqref{eq:hessianNoiseExp} that determines the effect of the perturbation on task error. Its contribution is:
\begin{align*}
\mathbb{E} [ \langle \ee, \nabla^2 \F[\w(0)] \ee \rangle]  &= \|\ee\|_2^2 \mathbb{E} [ \langle \he, \nabla^2 \F[\w(0)] \he \rangle]  \\
&=  \|\ee\|_2^2 \frac{Tr(\nabla^2 \F[\w(0)])}{N}.
\end{align*}

So the effect of random perturbations on learning grows with the ratio of $Tr(\nabla^2 \F[\w(0)])$ to the number of synapses, $N$. \eqref{eq:gdeDecomp} tells us explicitly how local task difficulty (and thus expected learning rate and steady-state performance) can be modified by changing the size of a network, provided the size change leaves $\GG^1[\w(0)]$ and $\frac{Tr(\nabla^2 \F[\w(0)])}{2 \| \nabla \F[\w(0)] \|_2 }$ unchanged. For different network architectures there are many possible ways of adding neurons and connections while satisfying these constraints. This explains why the naive size increases in Figure \ref{fig:2}C generically increased learning performance, and provides a general explanation for enhanced learning performance in larger networks.

\begin{figure}[t]
\centering
\includegraphics[width=1\linewidth]{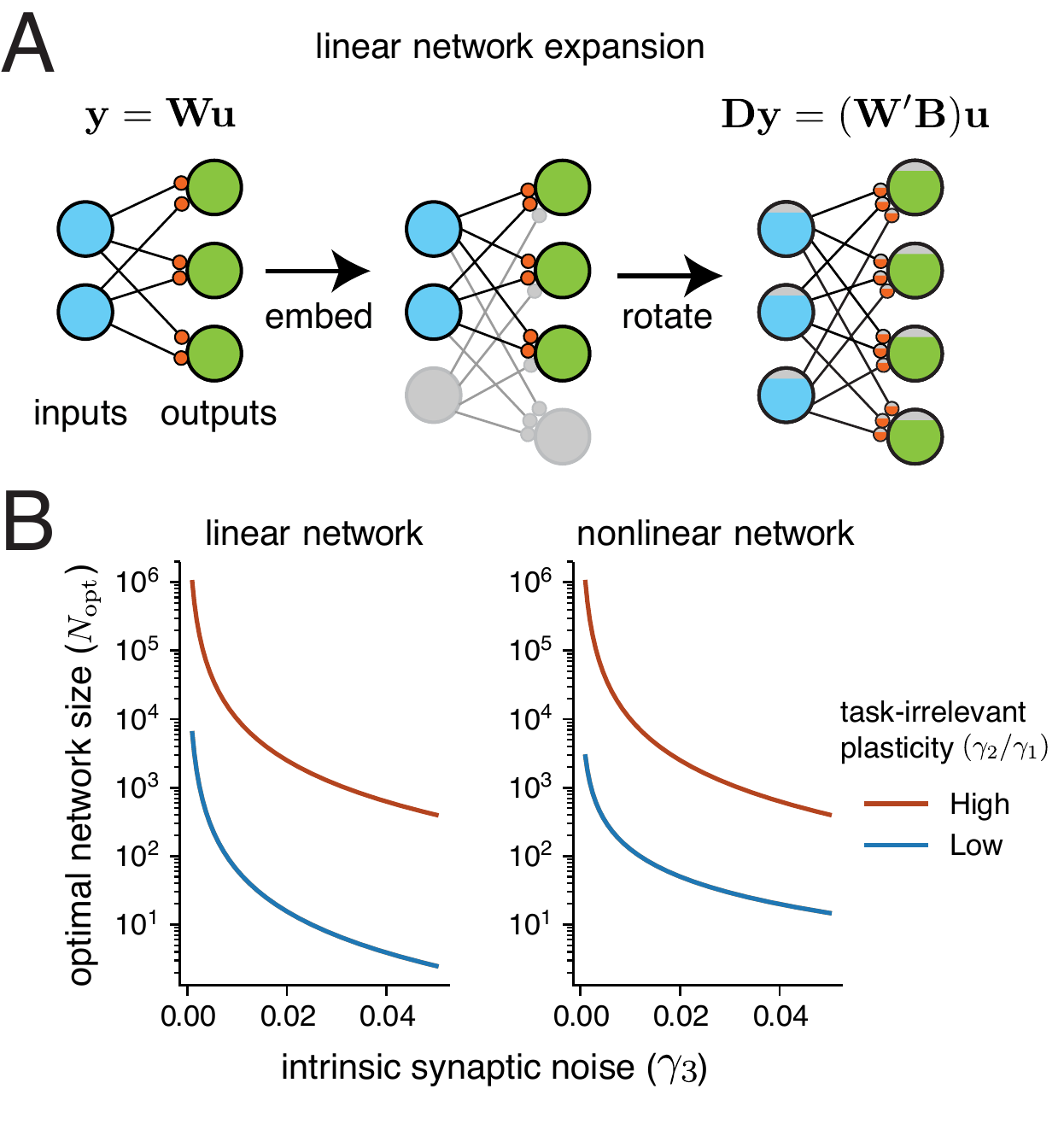} 
\caption{\label{fig:5} Optimal network size for linear and nonlinear networks in the presence of intrinsic synaptic noise. \textbf{A}: network expansion for a linear network, given by a embedding into a larger network, followed by a rotation of the weight matrix. This corresponds to transforming inputs $\mathbf{u}$ by a projection $\mathbf{B}$ and outputs $\mathbf{y}$ by a semiorthogonal mapping  $\mathbf{D}$.
\textbf{B}: Plots show the dependence of $N_\text{opt}$ in linear and nonlinear networks using equations \eqref{eq:optimalNlin} and \eqref{eq:optimalNnl}. In both cases the learning rule has $\gamma_1 = 0.1$. Low task-irrelevant plasticity corresponds to $\gamma_2 = 0.01$, while task-irrelevant plasticity corresponds to $\gamma_2 = 1$.
}
\end{figure}

\subsection*{Network expansions that increase learning performance}
We next give detailed examples of network expansions that increase learning rate and use the theory developed so far to compute the optimal size of a network when intrinsic noise is present. We first provide an analysis in the case of a linear network, which offers useful insight into the more general nonlinear case.

Consider a linear network (i.e. a linear map, as shown in Figure \ref{fig:5}A that transforms any input $u$ into an output $y = Wu$ for a matrix $W \in \mathbb{R}^{oi}$ of synaptic weights. 
 The input-dependent error of the network is taken as a simple mean square error:
\begin{align*}
\F[W] = \int \tF[\tW,\tu]  \mathbb{P}(u) \ du=  \int \| y^*(u) - Wu \|_2^2 \mathbb{P}(u) \ du.
\end{align*}
where the input vectors are drawn from some distribution $\mathbb{P}(u)$ (e.g. a Gaussian) and $y^*(u)$ is a target output generated by a linear mapping of the same rank and dimension.

We next embed this network in a larger network with $c_1i$ inputs, $c_2o$ outputs and a synaptic weight matrix $\tW$, for some integers $c_1$, $c_2 > 1$. We will define the total number of weights as $\tilde{N} = c_1ic_2o$.
We take the transformation $\tu = Bu \in \mathbb{R}^{c_1i}$, where $B \in \mathbb{R}^{c_1i \times i}$ is an arbitrary semi-orthogonal matrix. (i.e. it satisfies $B^TB = \mathbb{I}_i$). Geometrically, $B$ therefore represents the composition of a projection into the higher dimensional space $\mathbb{R}^{c_1i}$ with a rotation. Note that this is an invertible mapping: if $\tu = Bu$ then $B^T\tu = u$. Similarly, we can take $\ty^*(\tu) = Dy^*(u) \in \mathbb{R}^{c_2o}$, where $D^TD = \mathbb{I}_o$. This is illustrated in Figure \ref{fig:5}A. 

The expanded neural network with weights $\tW \in \mathbb{R}^{c_1o \times c_2i}$ has to learn the same mapping as the original, but with respect to the higher dimensional inputs. So the network receives inputs $\tu \in B\mathcal{U}$, and transforms them to outputs $\ty = \tW \tu$, with input-dependent error
\begin{align}
\tF[\tW,\tu] &= \|\ty^*(\tu) - \tW\tu \|_2^2 \nonumber \\
&= \|Dy^*(u) - \tW Bu \|_2^2 \nonumber \\
&= \|y^*(u) - D^T\tW Bu\|_2^2 \label{eq:linBigError}
\end{align}


For some weight configuration $\tW$ in an expanded network, \eqref{eq:linBigError} tells us that if these weights are related to the original network weights by $W = D^T \tW B \in \mathbb{R}^{oi}$, then we have:
\begin{align*}
\tF[\tW] = F[W].
\end{align*}

 If we rewrite the weight matrices $\tW$ and $W$ as vectors $\mathbf{w}' \in \mathbb{R}^{\tilde{N}}$ and $\w \in \mathbb{R}^N$, then $\w = H\mathbf{w}'$ for some $H$ with $HH^T = \mathbb{I}_N$. We can then apply the chain rule to relate gradients and hessians in the original and expanded networks:
\begin{align*}
&H \nabla \tF[\w'] = \nabla \F[\w], \\
&H \nabla^2 \tF[\w'] H^T = \nabla^2 \F[\w]. 
\end{align*}

Semi-orthogonality of $H$ implies that it has $N$ singular values with value one, and $\tilde{N}-N$ singular values with value zero. We take the heuristic that $\w'$ should project approximately equally onto each of the associated singular vectors. This is reasonable given that we took a random choice of $\w'$. This implies
\begin{align}
&\frac{\| \nabla \tF[\w' \|_2}{\|\nabla \F[\w]\|_2} \approx \frac{\tilde{N}}{N} = c_1c_2 \label{eq:linGradNorm}
\end{align}

Meanwhile the quadratic error function implies $\nabla^2 \tF[\w]$ is constant. The trace of this matrix can thus be calculated explicitly as
\begin{align}
Tr(\nabla^2\tF[\w']) = (c_2o)^2 \int_{u \in \mathcal{U}} \|u\|_2^4 \ \  \mathbb{P}(u). \label{eq:linTrace}
\end{align}

We will also assume that  $\nabla \htF[\w']$ (which is a normalised vector) projects approximately equally onto the different eigenvectors of $\nabla^2 \tF[\w']$. The latter is constant, whereas the former is a linear function of the (randomly chosen) $\tW$, which justifies this assumption. In this case, \eqref{eq:linTrace} implies
\begin{align}
\begin{split}
\nabla \htF[\w']^T \nabla^2 \tF[\w'] &\nabla \htF[\w'] \approx \\ &(c_2)^2 \nabla \hF[\w]^T \nabla^2 \F[\w] \nabla \hF[\w]. \label{eq:linGnum}
\end{split}
\end{align}

Bringing together equations \eqref{eq:linGradNorm}, \eqref{eq:linGnum}, and the formula \eqref{eq:GG1} for $\GG^1$, we see that
\begin{subequations} \label{eq:linearInvs}
\begin{align*}
\tGG^1[\w'] \approx \frac{(c_2)^2}{c_1c_2} \GG^1[\w] = \frac{c_2}{c_1} \GG^1[\w],
\end{align*}

similarly
\begin{align*}
\frac{Tr(\nabla^2\tF[\w'])}{\| \nabla \tF[\w'] \|_2} \approx  \frac{c_2}{c_1}\frac{Tr(\nabla^2\F[{\w}]) }{\| \nabla \F[\w] \|_2}.
\end{align*}
\end{subequations} 

Equations \eqref{eq:linearInvs} imply that we can alter $N$ as an independent parameter in the equations for local task difficulty \eqref{eq:gdeDecomp} and speed of weight change \eqref{eq:expGamma}, as long as the ratio $\frac{c_2}{c_1}$ is preserved. Indeed this allows us to optimise the steady state error of the network by changing $N$. To see how, recall that
\begin{align*}
\mathbb{E}[k] & =  \frac{ -\|\nabla \F[\w(0)]\|_2}{\F[\w(0)]}  \Big[ {-\gamma_1} + \dta \Big] + \mathcal{O}(T ^2), \\
\text{where } \dta &=  \GG[\hdw] \|\dw\|^2_2T,
\end{align*}

with $\mathcal{O}(T ^2) \equiv 0$ for quadratic error. Suppose the network has reached steady state error, i.e  $\mathbb{E}[k] = 0$. If we decreased $\dta$, then $\mathbb{E}[k]$ would also decrease, and the network would learn further. 
 Therefore, to derive the optimal $N^*$, we should minimise the expression for $\dta$ in $N$. We differentiate $\dta$ in $N$, and note that stationary points satisfy the equation:
\begin{align*}
&N^2 \left[ \gamma_1^2 CT + \gamma_3^2 \right] =
\frac{T^2\gamma_2^2}{\gamma_3^2} \big( \gamma_1^2 + \gamma_2^2) 
\\
 & \text{where } C = \frac{ \langle \nabla \hF[\w(0)], \nabla^2 \F[\w(0)] \nabla \hF[\w(0)] \Big\rangle }{Tr(\nabla^2 \F[\w(0)])}.
 \end{align*}
For $\gamma_2 \neq 0$, this implies the existence of two stationary points differing only in sign. Since $\lim_{N \to \infty} \dta = \infty$, and $\lim_{N \to -\infty} \dta = -\infty$, the positive stationary point is necessarily a global minimum of $\dta$. So this stationary point defines $N^*$. We have
  \begin{align}
 N_{\text{opt}}= \frac{T\gamma_2}{\gamma_3} \sqrt{ 
 \frac{1 + \frac{\gamma_2^2}{\gamma_1^2}}
 	{ CT + \frac{\gamma_3^2}{\gamma_1^2} }
 }. \label{eq:optimalNlin}
   \end{align}
   
\begin{figure}[t]
\centering
\includegraphics[width=0.8\linewidth]{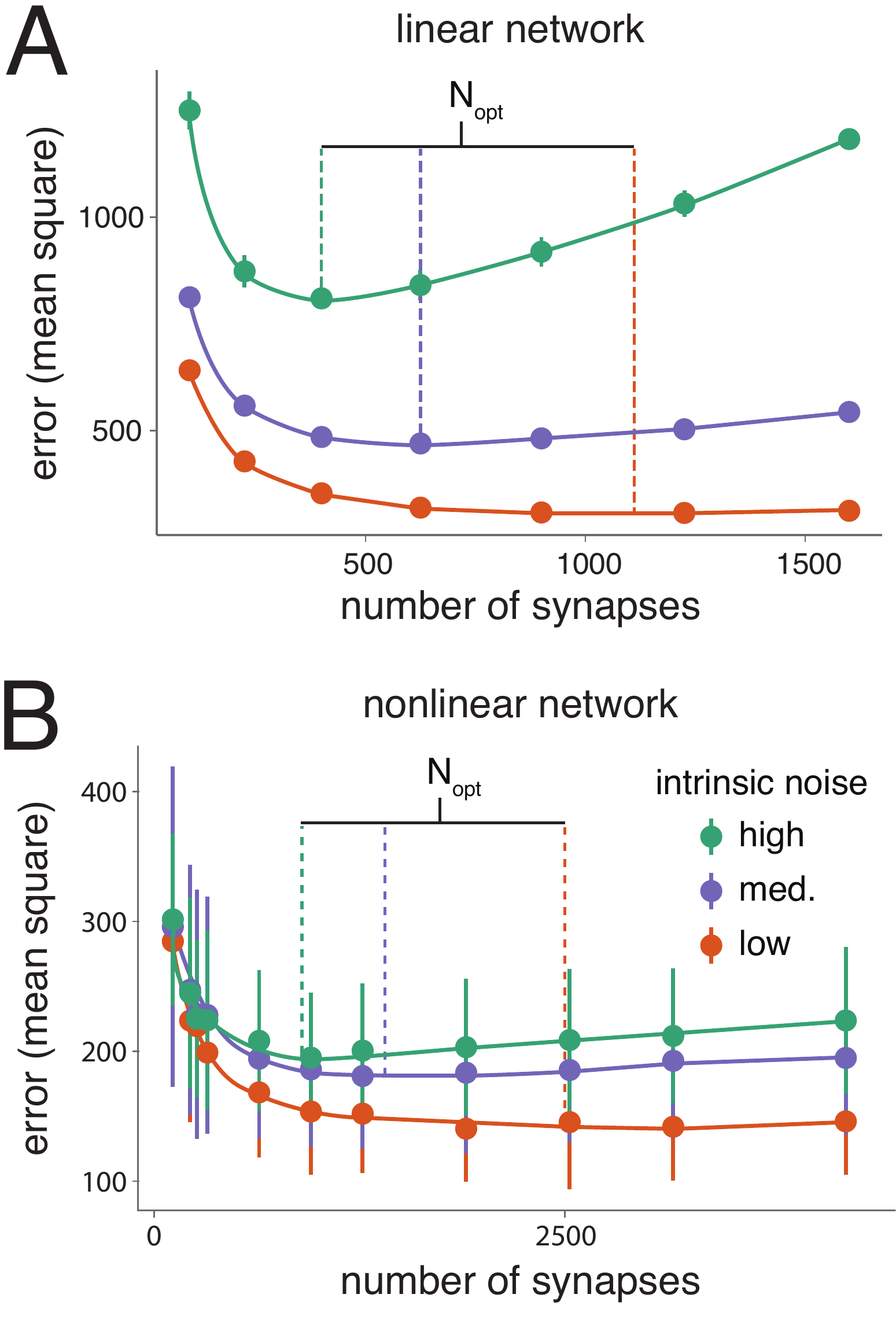} 
\caption{\label{fig:6} Testing analytic prediction of optimal network size for linear and nonlinear networks. Linear (A) and nonlinear (B) networks of different sizes are trained for $1500$ learning cycles of length $T=1$. Mean steady state error over $12$ repeats is plotted against network size. Coloured lines represent \emph{a priori} predicted optimal network sizes using equations \eqref{eq:optimalNlin} and \eqref{eq:optimalNnl} for the linear and nonlinear examples respectively.
\textbf{A}: Linear networks all have a $2:1$ ratio of inputs to outputs. On each repeat, networks of all considered sizes learn the same mapping, embedded in the appropriate input/output dimension (as detailed in {\em SI section `Learning in a Linear Network'}). The learning rule uses $\bg = [0.07,1,0.03]$ (high intrinsic noise), $\bg =  [0.06,1,0.04]$ (medium intrinsic noise) and $\bg  = [0.05,1,0.05]$ (low intrinsic noise). 
\textbf{B}: Nonlinear networks have sigmoidal nonlinearities at each neuron and a single hidden layer (see Methods). All networks have ten input and ten output neurons, and learn the same task. The number of neurons in the hidden layer is varied from $5$ to $120$. The learning rules all use $\bg_1 = 0.04$ and $\bg_2 = 1.5$. The value of $\bg_3$ is set respectively at $0.03$, $0.04$, and $0.05$, in the low, medium, and high intrinsic noise cases.
}
\end{figure}

Note that $C$ is unknown in general because it depends on the weight configuration of the network. However, we can take the heuristic $C \approx \frac{1}{N_{\text{opt}}}$, since if the gradient $\nabla \hF[\w(0)]$ is uncorrelated with the Hessian $\nabla^2 \hF[\w(0)]$, then it would project equally onto each of the eigenvalues of the latter, and thus the denominator of $C$ would be the mean eigenvalue, i.e. $\frac{Tr(\nabla^2 \hF[\w(0)])}{N}$. This results in an approximate expression for the optimal network size:
\begin{align}
 N_{\text{opt}}\approx \frac{T\gamma_2}{\gamma_3} \sqrt{ 
 \frac{1 + \frac{\gamma_2^2}{\gamma_1^2}}
 	{ \frac{T}{N_{\text{opt}}} + \frac{\gamma_3^2}{\gamma_1^2} }
 }, \label{eq:optimalNlin}
   \end{align}
   which can be solved for $N_{\text{opt}}$ algebraically. Note that the optimal $N_{\text{opt}}$ loses any dependence on the specific value of $\frac{Tr(\nabla^2 \F[\w(0)])}{2 \| \nabla \F[\w(0)] \|_2 }$. Our formula is verified numerically in Figure \ref{fig:6}A, by evaluating the learning performance of transformed neural networks of different sizes, with different $\gamma_i$ values.

This estimate of the optimal network size is plotted in Figure \ref{fig:5}B, which shows the dependence on intrinsic synaptic noise levels. As noise decreases to zero, we see that the optimal network size grows arbitrarily. In addition, the optimal network size is smaller for a lower amount of task-irrelevant plasticity (i.e. a `better' learning rule). We validate the optimal network size estimate in Figure \ref{fig:6}A with numerical simulations.

We next consider nonlinear multilayer, feedforward networks. Again, we use the student-teacher framework to generate learning tasks. We will consider learning performance of a nominal and expanded network, both with $l$ layers, and both using the same learning rule. The only difference between the two networks will be the larger number of neurons in each hidden layer of the expanded network. Unlike the linear case, this size expansion will modify some factors in the learning rate equation. Nevertheless, we can use our theory to predict an optimal number of synapses (and consequently optimal hidden layer sizes) for the transformed network. As before, this size will depend on the learning rule used by the networks, which is defined by levels of task-relevant plasticity, task-irrelevant plasticity, per-synapse white-noise intensity, and frequency of task error feedback. Our predictions are validated in simulations in Figure \ref{fig:6}B.

We first describe the nominal network architecture. Given a vector $h^{(k-1)}$ of neural activities at layer $k-1$, the neural activity at layer $h^{(k)}$ is
\begin{align*}
h^{(k)} = \underline{\sigma}(W^{(k)}h^{(k-1)}).
\end{align*}
Here, $W^{(k)}$ is the matrix of synaptic weights at the $k^{th}$ layer. The concatenation of the synaptic weight matrices across all layers is denoted $W$, and has $N$ elements. We will interchangeably denote it as a vector $\w \in \mathbb{R}^N$. The function $\sigma$ passes its arguments elementwise through some nonlinearity. The first layer of neurons receives an input vector $u$ in place of neural activities $h^{(0)}$. The output $y(W,u)$ is defined as neural activity at the final hidden layer.

For any given state $\w$ of the nominal network, we can construct a state $\phi(\w)$ of the expanded network with the same input-output properties, i.e. $y(\w,u) = y'(\phi(\w),u)$, where $y'$ denotes expanded network output. We do this by setting synaptic weights of the added neurons in the expanded network to zero, so the neurons do not contribute at state $\phi(\w)$. Nevertheless the extra neurons can affect expanded network behaviour because once they are perturbed by the learning rule they contribute to the error gradient and higher derivatives. 

Suppose the nominal network is at state $\w(0)$, and a learning rule picks the direction $\dw$ for weight change over the time interval $[0,T]$. This direction will have some local task difficulty $\GG[\hdw]$. If we map the state $\w(0)$ and the direction $\dw$ to the transformed network via the transformation $\phi$, then we can estimate local task difficulty $\GG'[\phi(\hdw)]$ of the transformed network (see {\em SI section `Learning in a Nonlinear, Feedforward Network' } for additional detail). We get:
\begin{align}
\mathbb{E}[\tGG[\phi(\hdw)]] \approx \frac{N}{\tilde{N}}\gamma_1^2 \GG^1[\w(0)] +  \frac{Tr(\nabla^2 \F[\w(0)])}{2 \| \nabla \F[\w(0)] \|_2 } \left[  \frac{\gamma_2^2  }{\tilde{N}} +    \frac{\gamma_3^2  }{T} \right]. \label{eq:nlGchange}
\end{align}
We can use \eqref{eq:nlGchange} to minimise
\begin{align*}
\dta &=  \GG'[\hdw] \|\dw\|^2_2T
\end{align*}
in $\tilde{N}$, the number of synapses. This gives an optimal size of the transformed network for minimising steady state task performance (validated in Figure \ref{fig:6}B):
\begin{align}
N_{\text{opt}} 
&\approx 
\frac{T\gamma_2}{\gamma_3^2} 
 \sqrt{
\Big(1 + \frac{\gamma_1^2N}{\gamma_2^2N_{\text{opt}}}\big)\big(\gamma_1^2 + \gamma_2^2 \big) 
}. \label{eq:optimalNnl}
\end{align}
Note the dependence of the optimal network size, $N_{\text{opt}}$, on $N$, which is the number of weights in the nominal and the teacher networks. The teacher networks can generate arbitrary nonlinear mappings whose complexity grows with $N$. In this way the above formula (Equation \ref{eq:optimalNnl}) reflects the {\em intrinsic difficulty} of the task.


\section*{Discussion}
It is difficult to disentangle the physiological and evolutionary factors that determine the size of a brain circuit \cite{chittka2009bigger,herculano2012remarkable,shepherd2005geometric}. Previous studies focused on the energetic cost of sustaining large numbers of neurons and connecting them efficiently \cite{tomasi2013energetic, herculano2012remarkable, shulman2004energetic, shepherd2005geometric}. Given the significant costs associated with large circuits \cite{attwell2001energy}, it is clear that some benefit must offset these costs, but it is currently unclear whether other inherent tradeoffs constrain network size. We showed under broad assumptions that there is an upper limit to the learning performance of a network which depends on its size and the intrinsic reliability of synapses.

The neural circuits in animals with large brains were presumably shaped on an evolutionary timescale by gradual addition of neurons and connections. Expanding a small neural circuit into a larger one can increase its dynamical repertoire, allowing it to generate more complex behaviours \cite{hinton2006reducing, tishby2015deep}. Less obviously, as we show here, circuit expansion can also allow a network to learn simpler tasks more quickly and to greater precision.

By directly analysing the influence of synaptic weight configurations on task error we derived a quantity we called `local task difficulty', that determines how easily an arbitrary network can learn. We found that local task difficulty always depends implicitly on the number of neurons, and can therefore be decreased by adding neurons according to relatively unrestrictive constraints. In simple terms, adding redundancy flattens out the mapping between synaptic weights and task error, reducing the local task difficulty on average. This flattening makes learning faster, and steady-state task error lower because the resulting error surface is less contorted and easier to descend using local task error information.

As an analogy, imagine hiking to the base of a mountain without a map, and doing so using intermittent estimates of the slope underfoot. A more even slope will be easier to descend because slope estimates will remain consistent. An undulating slope will be harder to descend because the direction of descent necessarily changes with location. Now consider the same hike in a heavy fog at dusk. The undulating slope will become far harder to descend. However, if it were possible to somehow smooth out the undulations (that is, reduce local task difficulty), the same hike would progress more efficiently. This analogy shows why larger neural circuits are able to achieve better learning performance in a given task where error information is corrupted.


In specific examples we show that adding neurons to intermediate layers of a multilayer, feedforward network, increases the magnitude of the slope (gradient) of the task error function relative to its curvature. From this we provide a template for scaling up network architecture such that both quantities increase approximately equally. This provides novel hypotheses for the organising principles in biological circuits which, among other things, predicts a prevalence of apparently redundant connections in networks that need to learn new tasks quickly and to high accuracy. Recent experimental observations reveal such apparently redundant connections in a number of brain areas across species \cite{druckmann2014structured,eichler2017complete,bartol2015nanoconnectomic,bloss2018single}.

Even if neurons are added to a network in a way that obeys the architectural constraints we derive, intrinsic synaptic noise eventually defeats the benefits conferred to learning. All synapses are subject to noisy fluctuations due to their molecular makeup \cite{ziv2017synaptic, mongillo2017intrinsic, minerbi2009long, attardo2015impermanence, puro1977synapse, faisal2008noise}. These sources of noise are distinct from shared noise in a feedback signal that is used in learning. Such independent noise sources accumulate as a network grows in size, outcompeting the benefit of size on learning performance. An immediate consequence is an optimal network size for a given task and level of synaptic noise.

Furthermore, our results show that different noise sources in nervous systems impact learning in qualitatively different ways. Noise in the learning rule as well as external noise in the task error, which may arise from sensory noise or fluctuations in the task, can be overcome in a larger circuit. On the other hand, the impact of intrinsic noise in the synaptic connections only worsens as network size grows.

Our analysis allowed us to predict the optimal size of a network in theoretical learning tasks where where we can specify the levels of noise in the learning rule and in synapses. Figure \ref{fig:5} shows that the optimal network size decreases rapidly as the intrinsic noise in synapses increases. We speculate that the emergence of large neural circuits therefore depended on evolutionary modifications to synapses that reduce intrinsic noise. An intriguing and challenging goal for future work would be to infer noise parameters in synapses across different nervous systems and test whether overall network size obeys the relationships our theory predicts.

\subsection*{Author Affiliations}

[a] Department of Engineering, University of Cambridge, Trumpington Street, CB21PZ.

\section{Methods}

We test our predictions in simulations of neural network models. This section describes the details of the models used. Code is available on request.

\subsection{Network architectures}
All tested neural networks have fully-connected feedforward architectures, to enable exact gradient computation. The simplest considered network is linear. Given an input $\uu \in \mathbb{R}^i$, this gives an output of the form
\begin{align*}
\uy(\uu) = W\uu \in \mathbb{R}^o,
\end{align*}
where $W \in \mathbb{R}^{oi}$ is a matrix of synaptic weights.

More commonly we consider networks with nonlinearities and hidden layers. In the main text, we refer to these as nonlinear, feedforward networks. Each neuron in these networks passes inputs through a sigmoidal nonlinearity $\sigma: \mathbb{R} \to \mathbb{R}$ of the form
\begin{align*}
\sigma(x) = \frac{1}{1 + \exp(-x)}.
\end{align*}
We use the notation $\underline{\sigma}:\mathbb{R}^m \to \mathbb{R}^m$ to represent the elementwise application of $\sigma$ to a vector of arbitrary length $m \in \mathbb{N}$. Let us denote neural activity at the $k^{th}$ layer via a vector $\uh^k$. If $k>1$, then
\begin{align*}
\uh^k = \underline{\sigma}(W^k\tilde{\uh}^{k-1}),
\end{align*}
where $W^k$ is a matrix of synaptic weights and $\tilde{\uh}^{k-1}$ is the concatenation of the vector $\uh^{k-1}$ with the scalar $-1$. This scalar is known as a bias neuron, and allows for nonzero $\uh^{k}$ even when each component of $\uh^{k-1}$ is zero. Meanwhile for $k=1$, we have
\begin{align*}
\uh^1 = \underline{\sigma}(W^1\uu),
\end{align*}
where $\uu \in \mathbb{R}^i$ is a vector of inputs to the network. Let $F$ be the final hidden layer of the network. Then network output $\uy \in \mathbb{R}^o$ is taken as
\begin{align*}
\uy =  \underline{\sigma}(W^k\tilde{\uh}^{F}),
\end{align*}
which will subsequently be denoted $\uy(\w,\uu)$. Here $\w$ is a vector that concatenates all entries of each weight matrix in the network. 

\subsection{Details of Learning Tasks}

For all training tasks used in simulation, we generate a random fixed mapping $\uy^*(\uu)$, along with a set of inputs $\mathcal{U}$, consisting of $1000$ elements. The mean square error of the network is given as
\begin{align*}
\sum_{u \in \mathcal{U}} \|\uy^*(\uu) - \uy(\w, \uu)\|_2^2.
\end{align*}
This corresponds to the error function $\F[\w]$ used in the main text.

We often consider a family of nonlinear neural networks learning the same task (i.e. the same mapping $\uy^*(\uu)$, using the same set $\mathcal{U}$ of inputs). Each network in the family has the same number of inputs, outputs, and hidden layers, but the number of hidden layer neurons varies. In this case, we must ensure \emph{a priori} that each neural network in the family is capable of learning the task perfectly, i.e. exactly recreating the mapping $\uy^*(\uu)$. 
To do this, we create a `teacher' network with the same architecture as the smallest network in the family (i.e. the network with the smallest number of hidden layer neurons). The synaptic weights of the teacher network are initialised randomly and then fixed, to generate a mapping $\uy^*(\uu)$ of the form described previously. Specifically, they are generated from a uniform distribution centred at $0$. The support of the distribution is set at $[-a,a]$, where $a$ is chosen such that the standard deviation of the weights is $\frac{1}{\sqrt{i}}$, where $i$ is the number of inputs the network receives. The weights of the smallest network in the family are randomly initialised from the same distribution. The weights of larger networks in the family are initialised such that the network's input-output behaviour exactly corresponds to the smallest network in the family. This is achieved by successively adding neurons to the smallest network in such a way that each addition does not change the input-output properties of the network. In this way, the initial task errors $\F[\w]$ of all networks in the family are identical. 

\subsection{Network Training}

All training protocols share the same basic template: they combine task relevant plasticity (i.e. gradient information), task-irrelevant plasticity processes, and a white-noise process at each synapse. The proportions of each of these three factors are respectively represented by parameters $\gamma_1$, $\gamma_2$, and $\gamma_3$. Meanwhile, the direction of plasticity changes upon reception of sensory feedback, which happens every $T$ units of time. We refer to each time interval  of length $T$ as a single learning cycle. The exact values of the $\gamma$ components will fluctuate between learning cycles (detailed subsequently). We can set a vector $\bg$ of their expected values. The synaptic weight vector is then updated according to the following formula:
\begin{align*}
\w_{t+T} = \w_t - T \bg_1 \nabla \hF[\w_t] + T \bg_2 \he_2^t + \sqrt{NT} \bg_3 \he_3^t.
\end{align*}
$N$ represents the number of synaptic weights, while $\he_3^t$ is a Gaussian random variable, normalised to ensure $\| \he_3^t\|_2 = 1$. Thus $\sqrt{NT} \bg_3 \he_3^t$  represents per-synapse white noise. 
Meanwhile, $\he_2^t$ is the state of a normalised random process uncorrelated with the task. The dynamics of the unnormalised random process satisfy $\he_2^{t+1} = \sqrt{0.1} \he_2^t + \sqrt{0.9} \hat{\nu}$, where $\nu$ is a Gaussian random variable, normalised such that $\|\nu\|_2 = 1$. So  $T \bg_2 \he_2^t $ represents the effect of partially systematic, task-irrelevant plasticity processes. 

Since 
\begin{align*}
\mathbb{E}[ \langle \nabla \F[\w]  , \he_i^t \rangle = 0, \quad i \in \{1,2\}
\end{align*}
we have that
\begin{align*}
\mathbb{E}[\gamma_i] = \bg_i, \quad i \in \{1,2,3\}.
\end{align*}
However, on a given learning cycle, the task-irrelevant components $\he_2^t$ and $\he_3^t$ will have some nonzero correlation with the gradient of task error. Consequently, the values of $\gamma$ on this learning cycle will differ from those of $\bg$. For instance, if the task irrelevant components anti-correlate with the gradient, then $\gamma_1$ will increase at the expense of $\gamma_2$ and $\gamma_3$.


\emph{We thank Fulvio Forni, Rodrigo Echeveste and Aoife McMahon for careful readings of the manuscript; we thank Rodolphe Sepulchre and Stephen Boyd for helpful discussions. This work is supported by ERC Grant StG2016 - FLEXNEURO (716643).}



\bibliography{rdBib}

\begin{thebibliography}{10}

\bibitem{laughlin1998metabolic}
Laughlin SB, van Steveninck RRdR, Anderson JC (1998) The metabolic cost of
  neural information.
\newblock {\em Nature neuroscience} 1(1):36.

\bibitem{tomasi2013energetic}
Tomasi D, Wang GJ, Volkow ND (2013) Energetic cost of brain functional
  connectivity.
\newblock {\em Proceedings of the National Academy of Sciences}
  110(33):13642--13647.

\bibitem{attwell2001energy}
Attwell D, Laughlin SB (2001) An energy budget for signaling in the grey matter
  of the brain.
\newblock {\em Journal of Cerebral Blood Flow \& Metabolism} 21(10):1133--1145.

\bibitem{reader2002social}
Reader SM, Laland KN (2002) Social intelligence, innovation, and enhanced brain
  size in primates.
\newblock {\em Proceedings of the National Academy of Sciences}
  99(7):4436--4441.

\bibitem{sol2005big}
Sol D, Duncan RP, Blackburn TM, Cassey P, Lefebvre L (2005) Big brains,
  enhanced cognition, and response of birds to novel environments.
\newblock {\em Proceedings of the National Academy of Sciences}
  102(15):5460--5465.

\bibitem{joffe1997visual}
Joffe TH, Dunbar R (1997) Visual and socio--cognitive information processing in
  primate brain evolution.
\newblock {\em Proceedings of the Royal Society of London B: Biological
  Sciences} 264(1386):1303--1307.

\bibitem{maguire2000navigation}
Maguire EA, et~al. (2000) Navigation-related structural change in the
  hippocampi of taxi drivers.
\newblock {\em Proceedings of the National Academy of Sciences}
  97(8):4398--4403.

\bibitem{gaser2003brain}
Gaser C, Schlaug G (2003) Brain structures differ between musicians and
  non-musicians.
\newblock {\em Journal of Neuroscience} 23(27):9240--9245.

\bibitem{black1990learning}
Black JE, Isaacs KR, Anderson BJ, Alcantara AA, Greenough WT (1990) Learning
  causes synaptogenesis, whereas motor activity causes angiogenesis, in
  cerebellar cortex of adult rats.
\newblock {\em Proceedings of the National Academy of Sciences}
  87(14):5568--5572.

\bibitem{lawrence1998size}
Lawrence S, Giles CL, Tsoi AC (1998) What size neural network gives optimal
  generalization? convergence properties of backpropagation, (UMIACS),
  Technical report.

\bibitem{krizhevsky2012imagenet}
Krizhevsky A, Sutskever I, Hinton GE (2012) Imagenet classification with deep
  convolutional neural networks in {\em Advances in neural information
  processing systems}.
\newblock pp. 1097--1105.

\bibitem{huang2003learning}
Huang GB (2003) Learning capability and storage capacity of two-hidden-layer
  feedforward networks.
\newblock {\em IEEE Transactions on Neural Networks} 14(2):274--281.

\bibitem{takiyama2016maximization}
Takiyama K (2016) Maximization of learning speed due to neuronal redundancy in
  reinforcement learning.
\newblock {\em Journal of the Physical Society of Japan} 85(11):114801.

\bibitem{takiyama2012maximization}
Takiyama K, Okada M (2012) Maximization of learning speed in the motor cortex
  due to neuronal redundancy.
\newblock {\em PLoS computational biology} 8(1):e1002348.

\bibitem{saxe2013exact}
Saxe AM, McClelland JL, Ganguli S (2013) Exact solutions to the nonlinear
  dynamics of learning in deep linear neural networks.
\newblock {\em arXiv preprint arXiv:1312.6120}.

\bibitem{seung2003learning}
Seung HS (2003) Learning in spiking neural networks by reinforcement of
  stochastic synaptic transmission.
\newblock {\em Neuron} 40(6):1063--1073.

\bibitem{werfel2004learning}
Werfel J, Xie X, Seung HS (2004) Learning curves for stochastic gradient
  descent in linear feedforward networks in {\em Advances in neural information
  processing systems}.
\newblock pp. 1197--1204.

\bibitem{contractor2015altered}
Contractor A, Klyachko VA, Portera-Cailliau C (2015) Altered neuronal and
  circuit excitability in fragile x syndrome.
\newblock {\em Neuron} 87(4):699--715.

\bibitem{rinaldi2008hyper}
Rinaldi T, Perrodin C, Markram H (2008) Hyper-connectivity and hyper-plasticity
  in the medial prefrontal cortex in the valproic acid animal model of autism.
\newblock {\em Frontiers in neural circuits} 2:4.

\bibitem{casanova2006minicolumnar}
Casanova MF, et~al. (2006) Minicolumnar abnormalities in autism.
\newblock {\em Acta neuropathologica} 112(3):287.

\bibitem{amaral2008neuroanatomy}
Amaral DG, Schumann CM, Nordahl CW (2008) Neuroanatomy of autism.
\newblock {\em Trends in neurosciences} 31(3):137--145.

\bibitem{loewenstein2015predicting}
Loewenstein Y, Yanover U, Rumpel S (2015) Predicting the dynamics of network
  connectivity in the neocortex.
\newblock {\em Journal of Neuroscience} 35(36):12535--12544.

\bibitem{ziv2017synaptic}
Ziv NE, Brenner N (2017) Synaptic tenacity or lack thereof: Spontaneous
  remodeling of synapses.
\newblock {\em Trends in neurosciences}.

\bibitem{minerbi2009long}
Minerbi A, et~al. (2009) Long-term relationships between synaptic tenacity,
  synaptic remodeling, and network activity.
\newblock {\em PLoS biology} 7(6):e1000136.

\bibitem{puro1977synapse}
Puro DG, De~Mello FG, Nirenberg M (1977) Synapse turnover: the formation and
  termination of transient synapses.
\newblock {\em Proceedings of the National Academy of Sciences}
  74(11):4977--4981.

\bibitem{bloss2018single}
Bloss EB, et~al. (2018) Single excitatory axons form clustered synapses onto
  ca1 pyramidal cell dendrites.
\newblock {\em Nature neuroscience} 21(3):353.

\bibitem{bartol2015nanoconnectomic}
Bartol~Jr TM, et~al. (2015) Nanoconnectomic upper bound on the variability of
  synaptic plasticity.
\newblock {\em Elife} 4:e10778.

\bibitem{chittka2009bigger}
Chittka L, Niven J (2009) Are bigger brains better?
\newblock {\em Current biology} 19(21):R995--R1008.

\bibitem{herculano2012remarkable}
Herculano-Houzel S (2012) The remarkable, yet not extraordinary, human brain as
  a scaled-up primate brain and its associated cost.
\newblock {\em Proceedings of the National Academy of Sciences} 109(Supplement
  1):10661--10668.

\bibitem{shepherd2005geometric}
Shepherd GM, Stepanyants A, Bureau I, Chklovskii D, Svoboda K (2005) Geometric
  and functional organization of cortical circuits.
\newblock {\em Nature neuroscience} 8(6):782.

\bibitem{shulman2004energetic}
Shulman RG, Rothman DL, Behar KL, Hyder F (2004) Energetic basis of brain
  activity: implications for neuroimaging.
\newblock {\em Trends in neurosciences} 27(8):489--495.

\bibitem{hinton2006reducing}
Hinton GE, Salakhutdinov RR (2006) Reducing the dimensionality of data with
  neural networks.
\newblock {\em science} 313(5786):504--507.

\bibitem{tishby2015deep}
Tishby N, Zaslavsky N (2015) Deep learning and the information bottleneck
  principle in {\em Information Theory Workshop (ITW), 2015 IEEE}.
\newblock (IEEE), pp. 1--5.

\bibitem{druckmann2014structured}
Druckmann S, et~al. (2014) Structured synaptic connectivity between hippocampal
  regions.
\newblock {\em Neuron} 81(3):629--640.

\bibitem{eichler2017complete}
Eichler K, et~al. (2017) The complete connectome of a learning and memory
  centre in an insect brain.
\newblock {\em Nature} 548(7666):175.

\bibitem{mongillo2017intrinsic}
Mongillo G, Rumpel S, Loewenstein Y (2017) Intrinsic volatility of synaptic
  connections---a challenge to the synaptic trace theory of memory.
\newblock {\em Current opinion in neurobiology} 46:7--13.

\bibitem{attardo2015impermanence}
Attardo A, Fitzgerald JE, Schnitzer MJ (2015) Impermanence of dendritic spines
  in live adult ca1 hippocampus.
\newblock {\em Nature} 523(7562):592.

\bibitem{faisal2008noise}
Faisal AA, Selen LP, Wolpert DM (2008) Noise in the nervous system.
\newblock {\em Nature reviews neuroscience} 9(4):292.

\end{thebibliography}

\end{document}